\documentclass[%
 reprint,
 amsmath,amssymb,
 aps,prd,
]{revtex4-1}

\usepackage{graphicx}
\usepackage{dcolumn}
\usepackage{bm}

\begin{document}

\preprint{APS/123-QED}

\title{Ultrahigh-Energy Cosmic Ray Production
by Turbulence in Gamma-Ray Burst Jets and Cosmogenic Neutrinos}

\author{Katsuaki Asano}
 \email{asanok@icrr.u-tokyo.ac.jp}
\affiliation{%
Institute for Cosmic Ray Research, The University of Tokyo,
5-1-5 Kashiwanoha, Kashiwa, Chiba 277-8582, Japan
}%


\author{Peter M\'esz\'aros}
 \email{nnp@psu.edu}
\affiliation{%
Department of Astronomy \& Astrophysics; Department of Physics;
Center for Particle \& Gravitational Astrophysics;
Pennsylvania State University, University Park, PA 16802
}%

\date{\today}

\begin{abstract}
We propose a novel model to produce ultrahigh-energy cosmic rays (UHECRs) 
in gamma-ray burst jets.  After the prompt gamma-ray emission,
hydrodynamical turbulence is excited in the GRB jets at or before
the afterglow phase. The mildly relativistic turbulence stochastically 
accelerates protons. The acceleration rate is much slower than the usual 
first-order shock acceleration rate, but in this case it can be 
energy independent.
The resultant UHECR spectrum is so hard that the bulk energy
is concentrated in the highest energy range, resulting in a
moderate requirement for the typical cosmic-ray luminosity 
of $\sim 10^{53.5}~\mbox{erg}~\mbox{s}^{-1}$.
In this model, the secondary gamma-ray and neutrino emissions
initiated by photopion production are significantly suppressed.
Although the UHECR spectrum at injection shows a curved feature,
this does not conflict with the observed UHECR spectral shape.
The cosmogenic neutrino spectrum in the $10^{17}$--$10^{18}$ eV
range becomes distinctively hard in this model, which may be
verified by future observations.
\end{abstract}

\pacs{Valid PACS appear here}
\maketitle


\section{\label{sec:intro}Introduction}

The origin of ultrahigh-energy cosmic rays (UHECRs)
above the ankle energy ($\sim 10^{18.5}$ eV)
is a matter of ongoing discussions.
Although  jets in active galactic nuclei (AGN) are the 
most widely considered candidates for the UHECR sources \citep{rac93},
the observed degree of anisotropy in the arrival distribution
indicates a source density larger than $10^{-4}~\mbox{Mpc}^{-3}$
for a pure proton compositon ($10^{-6}~\mbox{Mpc}^{-3}$
for a pure iron composition) \citep{tak09,tak12},
which disfavors Fanaroff-Reily II galaxies and BL Lac objects.
Other types of relatively low-luminosity AGNs like Seyfert galaxies may 
not satisfy the luminosity requirement ($\gtrsim 10^{46}~\mbox{erg}~\mbox{s}^{-1}$
for protons) needed for UHECR acceleration \citep{nor95}.
Clusters of galaxies with strong accretion shocks
are also candidates of UHECR sources \citep{kan96,ino05}.
However, while the Telescope Array experiment reported a cluster of 
UHECR events \citep{abb14} in a $20^\circ$ radius (TA hot spot), there is 
no clear excess in the direction toward the nearby massive cluster, Virgo.

This situation motivates us to revisit gamma-ray bursts (GRBs) as UHECR sources 
\citep{vie95,wax95}, although the GRB hypothesis has been considered to have
disadvantages. The prompt emission of GRBs is believed to be emitted from
collimated ultrarelativistic outflows.  In most of the GRB UHECR models,
the internal shocks formed in the GRB outflows \citep{pir04,mes06}
are supposed to be the UHECR acceleration site. In this case, the shock 
accelerated particles (hereafter we assume UHECR to be protons) form a 
power-law number spectrum ($N(\varepsilon) \propto \varepsilon^{-p}$)
with a typical index of $p \sim 2$.
The observed GRB rate \footnote{A narrow (wide) average jet opening angle 
leads to a high (low) actual GRB rate but low (high) energy release per burst.
Thus, roughly speaking, the actual energy release rate from GRBs is 
independent of the jet opening angle.  Hereafter, we adopt the observed 
GRB rate with the isotropically equivalent energy release.}
is so low that the required cosmic-ray luminosity to agree with the observed
UHECR flux is 30--100 times the gamma-ray luminosity or more
\citep[see e.g.,][]{mur08,bae15}.  Such a high luminosity seems unfavorable 
in the light of the available energy budget.
Assuming a cosmic-ray luminosity much larger than the gamma-ray luminosity,
the secondary neutrino flux has been calculated by many authors
\citep[e.g., Refs.][]{gue04,mur08,bae15,bus15,bus16}.
However, the IceCube neutrino telescope has detected no significant 
high-energy neutrino emission associated with classical GRBs \citep{abb12,gao13,aar15-GRB}.
This severely constrains the UHECR luminosity in GRBs,
although a moderate value of the ratio of the UHECR to gamma-ray luminosity
($f_{\rm CR}\sim10$) is allowed \citep{He12,asa14}. In particular,
a GRB UHECR model with the moderate ratio $f_{\rm CR}=10$ can reproduce 
the observed UHECR flux, but only above $\sim 10^{20}$ eV \citep{asa14}.
Furthermore, while most of the previous studies of the GRB neutrino emission
associated with UHECR have omitted a discussion of the secondary gamma rays,
the required high UHECR luminosity must result in a spectral shape of the 
gamma rays which differs from the typically observed ones, as result of the 
hadronic cascade initiated by the collisions with gamma-ray photons
\citep{asa09,asa14,pet14}.

In this paper, we discuss a different scenario for the UHECR production in GRBs,
which may avoid the above-mentioned problems. Among the latter, the
major difficulty arises because the power-law spectrum of index $p=2$ with 
an exponential cutoff, which has been frequently assumed in the shock acceleration 
model, leads to a large energy fraction residing below the ankle energy.
On the other hand, if the spectral index were shallower than 2, then most 
of the cosmic-ray energy would be concentrated around the highest 
energy range. This could reduce the total proton energy budget,
putting the bulk of the UHECR energy above the ankle.
Such a hard spectrum would probably involve a different acceleration 
mechanism or acceleration site from those in the internal shock models.
Another requirement in the GRB UHECR model is the suppression
of the secondary gamma rays and neutrinos, the first being constrained
by Fermi and the second by IceCube observations. This suggests that
the UHECR acceleration site should be significantly outside the usual
photon emission site, in order to reduce the photopion production efficiency.
Such a setup could provide a convincing solution for avoiding an overly luminous 
gamma-ray/neutrino emission compatible with the required large UHECR luminosity,
unless the average bulk Lorentz factor of the GRB jets is $\gtrsim 1000$ \citep{asa09},
or the gamma-ray emission site is also located at a larger distance
than usually assumed \citep{pet14b,bus16}.
In some observational \citep{gui15,gui16}
and theoretical \citep[e.g., Ref.][]{bus16} studies, it has been argued that
multiple spectral components of the prompt $\gamma$-ray emission may arise from 
different emission sites or mechanisms. This encourages us to consider a 
different site for the UHECR acceleration.

An alternative model for the prompt gamma-ray emission is the
dissipative photosphere model, which has been discussed by Refs.
\citep{mes00,gia11,pee06,iok07,laz10,bel10,asa13}. In this model,
the photon emission site is at a distance $\sim 10^{10}-10^{13}$ cm 
from the central engine, leaving a large fraction of the bulk kinetic 
energy of the flow to be dissipated at a larger distance.
Numerical simulations of the deceleration of such relativistic
outflows \citep{duf13} show the development of a Rayleigh-Taylor instability 
at large distances $\gtrsim 10^{16}$ cm.
In such regions, stochastic acceleration via turbulence can accelerate UHECRs, 
and the photopion production efficiency will be significantly low.
The stochastic acceleration can yield a hard UHECR spectrum with $p<2$ 
\citep{sch84,par95,bec06,sta08}, having been discussed as a possible electron 
acceleration mechanism in AGN jets \citep{bot99,sch00,kat06,kak15}.
Stochastic acceleration of electrons via turbulence has also been discussed 
in connection with the mechanism of the GRB prompt gamma-ray emission
\citep{Byk96,der01,asa09b,mur12,asa15b}.
In particular, recent numerical simulations of the stochastic acceleration
and photon emission in AGN jets \citep{asa14b,asa15}
succeed in reproducing the wide-band spectra from radio to gamma rays
and the gamma-ray light curves, showing that stochastic acceleration
in relativistic outflows is an attractive option.

In this paper, we propose a UHECR production model in GRB outflows
based on the stochastic acceleration of protons via turbulence
at a large distance, well outside the photon emission site.
As will be shown, this optimized model can avoid both the problems of
an overly large UHECR loading and an overly luminous secondary gamma-ray/neutrino
emission.

\section{Stochastic Acceleration}

We consider turbulence excited in
a relativistic jet outflowing with the bulk Lorentz factor $\Gamma$.
As possible mechanisms to excite turbulence,
in addition to the Rayleigh-Taylor instability
in the decelerating outflow \citep{duf13}, there is
the Kelvin-Helmholtz instability in the shear flow
\citep[e.g., Refs.][]{Zha03,Miz09}
or at the boundary between
the jet and cocoon \citep{Mes01,Ram02}.
The Rayleigh-Taylor and Richtmyer-Meshkov instabilities
are also candidates for inducing
turbulence as radial modes \citep{mat13}.
The Rayleigh-Taylor and Kelvin-Helmoholtz instabilities
may be suppressed in the presence of a large-scale magnetic field
\citep{jun95,jun96,sto07,ran07},
depending on the orientation of the field.
However, since the magnetic energy is subdominant
compared to the hadronic energy in our model, as will be seen,
we can expect significant excitation of turbulence.
Another possible process is the internal shock
with density fluctuation, which induces the Richtmyer-Meshkov instability
\citep{ino11}.
The induced turbulence can scatter charged particles,
which causes the second-order Fermi acceleration \citep{ski75,sch84,ptu88,cho06}.
In addition, the turbulence may enhance the rate of magnetic reconnection
\citep{laz99}, which can also accelerate particles stochastically
\citep{hos12,kag13,guo14,guo15}.
Although there are many candidates, we do not specify the particular (magneto)hydrodynamical
instability that is responsible for particle acceleration.
We define an energy diffusion coefficient,
\begin{eqnarray}
D(\varepsilon) \equiv \frac{1}{2}
\left< \frac{\Delta \varepsilon \Delta \varepsilon}{dt} \right>,
\end{eqnarray}
which phenomenologically describes the stochastic acceleration
process of relativistic protons $\varepsilon \gg m_{\rm p}c^2$
in the jet comoving frame.

As the most optimistic model, we adopt the formula for the energy diffusion coefficient 
via magnetic field compressions associated with compressional waves \citep{Lyn14} as
\begin{eqnarray}
D(\varepsilon) \sim \varepsilon^2 \frac{v_{\rm W}^2}{c \ell}
\frac{\delta B^2}{B^2} \int_{1}^{\ell k_{\rm max}}
d(\ell k) (\ell k)^{1-q},
\label{EqLyn}
\end{eqnarray}
where $v_{\rm W}$ is the phase velocity of the turbulent wave,
and the turbulence is characterized by
the injection (longest) scale $\ell$ and
the index of the power spectrum $q$ as a function of
wave number $k$.
The case $D(\varepsilon) \propto \varepsilon^2$ is called
the hard sphere approximation.
Even if we consider the Alfv\'enic wave as a scatterer,
the turbulence index $q=2$ frequently seen in magnetohydrodynamic simulations
\citep[e.g., Refs.][]{kow10} results in the hard sphere approximation \citep{bla87}.
Note that the above formulation is based on the quasilinear theory
\citep{mel68}, which assumes $\delta B^2 \ll B^2$.
However, the self-generated magnetic fields in the jet may imply
$\delta B^2 \sim B^2$. In addition, the simulations in the work by \citet{Lyn14}
show a deviation from the quasilinear theory in the acceleration process.
We characterize the uncertainty in the acceleration process
related to $\delta B^2 \sim B^2$ and the unknown parameters 
$q$ and $k_{\rm max}$ in Eq. (\ref{EqLyn}) by a 
dimensionless factor $\zeta$ and write the diffusion coefficient as
\begin{eqnarray}
D(\varepsilon) \sim \varepsilon^2 \beta_{\rm W}^2 \zeta \frac{c}{\ell}.
\end{eqnarray}
In the turbulence excited by relativistic motions,
the phase velocity may be close to the
relativistic limit $\beta_{\rm W}^2=v_{\rm W}^2/c^2=1/3$.
The eddy scale $\ell$ should be shorter than the fluid-frame dynamical
scale $R/\Gamma$ at radius $R$ from the central engine.
We parametrize the eddy scale through a dimensionless parameter $\xi$ as
$\ell=\xi R/\Gamma \equiv 0.1 \xi_{0.1} R/\Gamma$.
Then, the diffusion coefficient is rewritten as
\begin{eqnarray}
D(\varepsilon) \sim 3 \varepsilon^2 \zeta \frac{c \xi_{0.1} \Gamma}{R}.
\label{dpp}
\end{eqnarray}
Hereafter, we denote $D(\varepsilon) \equiv K \varepsilon^2$
($K=3 c \zeta \xi_{0.1} \Gamma/R$).
The acceleration time scale $t_{\rm acc} \sim 1/K \gtrsim t_{\rm dyn}/(3 \zeta)$
($t_{\rm dyn}=R/(c \Gamma)$) is independent of the proton energy.
This implies that the mean free path is also energy independent.
A recent 3D MHD simulation \citep{por16} for pulsar wind nebulae
shows such a tendency as well.
The diffusion coefficient in
Eq. (\ref{dpp}) may be the most optimistic case for turbulence acceleration
and is the counterpart of the Bohm limit in the shock acceleration,
which has been frequently assumed in the UHECR acceleration in GRBs.

To be accelerated by the mechanism described above,
the Larmor radius is required to be shorter than the eddy size $\ell$.
The magnetic field in the jet frame is normalized by the photon energy density
as
\begin{eqnarray}
U_{\rm ph}=\frac{L_{\gamma}}{4 \pi c R^2 \Gamma^2},
\end{eqnarray}
where $L_{\gamma}$ is the isotropic-equivalent luminosity
of the GRB prompt emission.
Expressing the magnetic field with a dimensionless parameter $f_B$
as $B^2/(8 \pi)=f_B U_{\rm ph}$,
the maximum energy of protons for observers,
$\varepsilon_{\rm max}=\Gamma \ell e B/(1+z)$,
is written as
\begin{eqnarray}
\varepsilon_{\rm max}&=&\frac{\xi e}{(1+z)\Gamma}
\sqrt{\frac{2 f_B L_{\gamma}}{c}}  \label{emax} \\
&\simeq& 8.2 \times 10^{19} \xi_{0.1} (1+z)^{-1}
\Gamma_{300}^{-1} f_B^{1/2}
 L_{52}^{1/2}~\mbox{eV},
\label{emax2}
\end{eqnarray}
where $\Gamma=300 \Gamma_{300}$ and $L_{\gamma}=10^{52} L_{52}$
erg~$\mbox{s}^{-1}$.
Note that the maximum energy is independent of $R$.
Although the maximum energy is lower by a factor of $\xi$
than the value in the Bohm limit approximation for shock acceleration, 
the value in Eq. (\ref{emax2}) is significantly high to explain the 
observed UHECR spectrum. As mentioned, the UHECR acceleration site
is taken here to be well outside the gamma-ray emission radius.
Thus, the cooling effect due to photopion production can be neglected.
The time scale of the proton synchrotron cooling $t_{\rm syn}
=6 \pi (m_{\rm p}/m_{\rm e})^2
m_{\rm p}^2 c^3/(\sigma_{\rm T} B^2 \varepsilon)$
is also long enough,
\begin{eqnarray}
\frac{t_{\rm syn}}{t_{\rm dyn}} \simeq 1600
\varepsilon_{\rm obs,19}^{-1} \Gamma_{300}^{4} f_B^{-1} L_{52}^{-1}
R_{16},
\end{eqnarray}
where $\Gamma \varepsilon=10^{19} \varepsilon_{\rm obs,19}$ eV
and $R \equiv 10^{16} R_{16}$ cm.

The evolution of the total proton energy distribution $N(\varepsilon,t)$ in the 
jet frame is described by the Fokker-Planck equation \citep[e.g., Refs.][]{pet04,sta08}
as
\begin{eqnarray}
\frac{\partial N(\varepsilon,t)}{\partial t}
=&&\frac{\partial}{\partial \varepsilon}
\left[ D(\varepsilon) \frac{\partial N(\varepsilon,t)}{\partial \varepsilon}
\right]\nonumber \\
&&-\frac{\partial}{\partial \varepsilon}
\left[ \frac{2 D(\varepsilon)}{\varepsilon} N(\varepsilon,t)
\right]+\dot{N}_{\rm inj}(\varepsilon,t)
\label{eq:one},
\end{eqnarray}
where $\dot{N}_{\rm inj}(\varepsilon,t)$ is the injection term.
For simplicity, we assume that the jet is filled with turbulence
and the strong magnetic field, which implies small Larmor radii $<0.1 \xi_{0.1} R/\Gamma$,
efficiently confines particles in the dynamical time scale.
Hence, we have neglected the escape effect in Eq. (\ref{eq:one}),
though the escape effect can soften the cosmic-ray spectrum
\citep{bec06,sta08}.
If the coefficient $K$ can be approximated as being constant, we can 
obtain the Green's function for Eq. (\ref{eq:one}) as
derived by \citet{bec06},
\begin{equation}
N_{\rm G}(\varepsilon,t)
=\frac{N_0}{2 \varepsilon_0 \sqrt{\pi K t}}
\sqrt{\frac{\varepsilon}{\varepsilon_0}}
\exp{\left( -\frac{9}{4} K t -\frac{(\ln{\frac{\varepsilon}{\varepsilon_0}})^2}
{4 Kt} \right)},
\end{equation}
which corresponds to the solution of Eq. (\ref{eq:one})
for a prompt monoenergetic injection at $t=0$
with $\dot{N}_{\rm inj}(\varepsilon,t) \equiv N_0 \delta(\varepsilon-\varepsilon_0)
\delta(t)$.
For a constant injection with
$\dot{N}_{\rm inj}(\varepsilon) \equiv \dot{N}_0 \delta(\varepsilon-\varepsilon_0)$
for $t \geq 0$,
the solution is obtained by integrating the Green's function as
\begin{eqnarray}
N(\varepsilon,t)&=&\frac{\dot{N}_0}{N_0} \int_0^t dt' N_{\rm G}(\varepsilon,t') \\
&=&\frac{\dot{N}_0}{2 \varepsilon_0}
\sqrt{\frac{\varepsilon}{\varepsilon_0}}
\int_0^t dt'
\frac{\exp{\left( -\frac{9}{4} K t' -\frac{(\ln{\frac{\varepsilon}{\varepsilon_0}})^2}
{4 Kt'} \right)}}{\sqrt{\pi K t'}}. \nonumber \\ \label{integ}
\end{eqnarray}
The integral in Eq. (\ref{integ}) is analytically obtained
with the error function ${\rm erf}(x) \equiv (2/\sqrt{\pi}) \int_0^x dy \exp{(-y^2)}$
as
\begin{eqnarray}
N(\varepsilon,t)&=&\frac{\dot{N}_0}{6 K \varepsilon_0}
\sqrt{\frac{\varepsilon}{\varepsilon_0}} \nonumber \\
&&\times \left[ \exp{\left( -\frac{3}{2} \left|
\ln{\frac{\varepsilon}{\varepsilon_0}} \right| \right)}
(1+{\rm erf}(X_-)) \right. \nonumber \\
&& \left. +\exp{\left( \frac{3}{2} \left|
\ln{\frac{\varepsilon}{\varepsilon_0}} \right| \right)}
(-1+{\rm erf}(X_+)) \right],
\end{eqnarray}
where
\begin{eqnarray}
X_\pm \equiv \frac{3 K t \pm \left| \ln{\frac{\varepsilon}{\varepsilon_0}} \right|}
{2 \sqrt{Kt}}.
\end{eqnarray}
For $\varepsilon \geq \varepsilon_0$, the spectrum can be rewritten as
\begin{equation}
N(\varepsilon,t)=\frac{\dot{N}_0}{6 K \varepsilon}
\left[ 1+{\rm erf}(X_-)-\left( \frac{\varepsilon}{\varepsilon_0}
\right)^3{\rm erfc}(X_+) \right],
\label{eq:evol}
\end{equation}
where ${\rm erfc}(x) \equiv 1-{\rm erf}(x)$ is the complementary error function.
On the other hand, the distribution
for $\varepsilon \leq \varepsilon_0$ is approximated by
a steady solution:
\begin{eqnarray}
N(\varepsilon,t) \simeq \frac{\dot{N}_0}{3 K \varepsilon_0}
\left( \frac{\varepsilon}{\varepsilon_0}
\right)^2.
\end{eqnarray}

\begin{figure}[b]
\includegraphics[width=0.45\textwidth]{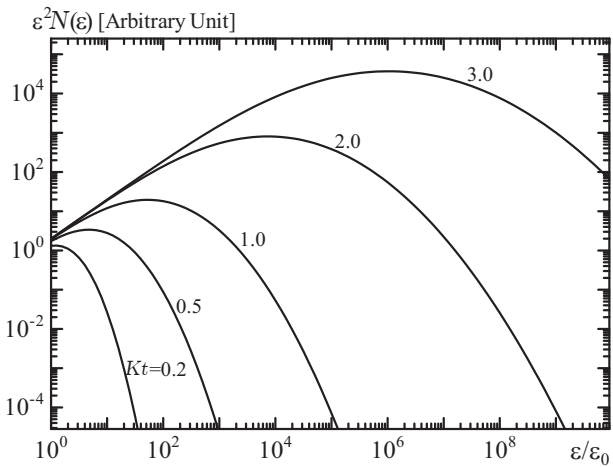}
\caption{\label{fig:evol} Evolution of the particle energy distribution
expressed by Eq. (\ref{eq:evol}).}
\end{figure}
The evolution of $N(\varepsilon,t)$ for constant $K$, $\varepsilon_0$,
and $\dot{N}_0$
is shown in Fig.~\ref{fig:evol}.
Since the acceleration time scale is independent of energy,
the peak energy in a $\varepsilon^2 N(\varepsilon,t)$ plot
is very sensitive to the duration $t$ of the acceleration and injection.
In this hard sphere model, a slight increase of the duration time
drastically boosts the spectral shape so that
it may be very difficult to determine the spectral shape
precisely for individual cases.

The total UHECR energy in the central engine frame is expressed as
\begin{eqnarray}
E(t) &=& \Gamma \int d \varepsilon \varepsilon N(\varepsilon,t) \\
&\equiv& \frac{\dot{N}_0 \Gamma}{6 K} \varepsilon_0 I(Kt).
\end{eqnarray}
Here, the dimensionless function $I(Kt)$ can be approximated as
\begin{eqnarray}
I(Kt) \simeq \left\{ \begin{array}{ll}
1.49 \exp{(4Kt)} & \mbox{for}~Kt \geq 1 \\
6.34 Kt & \mbox{for}~Kt \ll 1. \\
\end{array} \right.
\end{eqnarray}
In our optimistic model, during the dynamical time scale $t_{\rm dyn}=R/(c \Gamma)$,
the parameters $K$, $\varepsilon_0$, and $\dot{N}_0$ are assumed to be constant.
Just after the dynamical time scale, we shut down the acceleration and injection.
In this case, the value $K t_{\rm dyn}\sim 3\zeta \xi_{0.1}$ becomes independent of $R$.
Neglecting adiabatic cooling before escaping from the acceleration region,
the total energy of the UHECRs escaping from a GRB is
$E_{\rm CR}=E(t_{\rm dyn})$.
Hereafter, we normalize $E_{\rm CR}$ through the total GRB energy in photons
as $E_{\rm CR}=f_{\rm CR} E_{\gamma}$.
An empirical relation obtained by \citet{Ghi12} is
\begin{eqnarray}
E_{52}=0.56 L_{52}^{1.1},
\label{eq:ghirel}
\end{eqnarray}
where $E_{\gamma}=10^{52} E_{52}$ erg.
This is used to fix the normalization factor in Eq. (\ref{eq:evol})
as
\begin{eqnarray}
\frac{\dot{N}_0}{6K}=
\frac{f_{\rm CR} E_{\gamma}}{\Gamma \varepsilon_0 I(3\zeta\xi_{0.1})},
\label{eq:norm}
\end{eqnarray}
which is also independent of $R$.

The injection mechanism into the acceleration process is highly uncertain.
If the Kelvin-Helmholtz instability between the jet and the cocoon
is responsible for the stochastic acceleration,
the initial relative Lorentz factor $\sim \Gamma$ for the two layers may 
correspond to a typical random Lorentz factor of protons in the disturbed region.
We adopt an injection energy written as $\varepsilon_0=\Gamma m_{\rm p} c^2$.

\begin{figure}[t]
\includegraphics[width=0.45\textwidth]{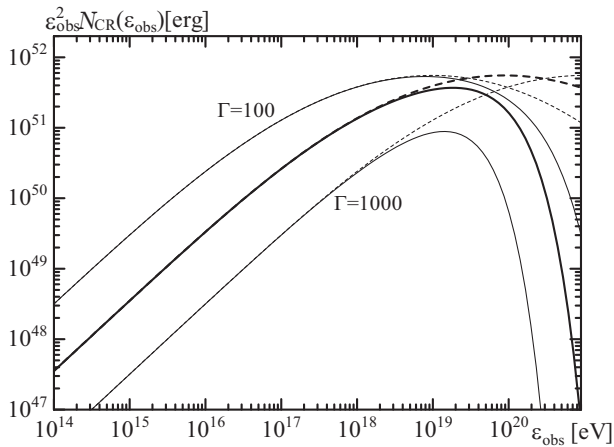}
\caption{\label{fig:ex} Model spectra of the UHECRs escaping from a GRB.
The thick lines are the spectrum for the parameters
$L_{52}=\Gamma_{300}=f_B=f_{\rm CR}=\xi_{0.1}=1$,
while the thin lines show the spectra with the same parameters
but for different values of $\Gamma$.
The dashed lines are the spectra neglecting the exponential cutoff
due to the maximum energy determined by the eddy size.}
\end{figure}

While the spectrum $N(\varepsilon,t_{\rm dyn})$ largely depends on the uncertain 
parameters $K t_{\rm dyn}$ and $\varepsilon_0$, these parameters have been roughly 
fixed as explained above. However, our model does not produce unbridled UHECR spectra,
because we have introduced a maximum energy defined in Eq. (\ref{emax2}).
Taking into account this maximum energy by simply introducing an exponential cutoff, 
we finally obtain the observer-frame spectrum of UHECRs from a GRB as
\begin{eqnarray}
N_{\rm CR}(\varepsilon_{\rm obs})
&=& \frac{1+z}{\Gamma}
N\left( \frac{(1+z) \varepsilon_{\rm obs}}{\Gamma},t_{\rm dyn} \right) \nonumber \\
&& \times \exp{\left( -\frac{\varepsilon_{\rm obs}}{\varepsilon_{\rm max}} \right)},
\label{eq:spec}
\end{eqnarray}
where $\varepsilon_{\rm obs}=\Gamma \varepsilon/(1+z)$.
In the above formula, protons are assumed to escape promptly in
an energy-independent manner,
and the cooling effects during propagation
in the intergalactic space are neglected.
The spectrum does not depend on $R$, and the model parameters for a single 
GRB are $L_{52}$, $\Gamma_{300}$, $f_B$, $f_{\rm CR}$, $\zeta$, and $\xi_{0.1}$.
Examples of the UHECR spectra expressed by Eq. (\ref{eq:spec})
are shown in Fig. \ref{fig:ex}.
The factor of the exponential cutoff significantly reduces
the actual energy of UHECRs compared to the prearranged value $E_{\rm CR}$.
Although we can adjust the energy of UHECRs including the cutoff effect,
we adopt the normalization factor in Eq. (\ref{eq:norm}) for simplicity.

\section{Average Spectrum per burst}

Hereafter, we will fix the parameters $f_B$, $\zeta$, and $\xi_{0.1}$ to be unity.
We adopt the GRB luminosity function obtained by \citet{wan10}, taking
the GRB rate per unit comoving volume per logarithmic interval of luminosity 
defined as $\phi(L_\gamma) {\cal R}_{\rm GRB}(z) d \log L_\gamma$. 
This luminosity function is written as
\begin{eqnarray}
\phi(L_\gamma)  \propto \left\{ \begin{array}{ll}
\left( \frac{L_\gamma}{L_*} \right)^{-0.17} & \mbox{for}~L_\gamma \leq L_* \\
\left( \frac{L_\gamma}{L_*} \right)^{-1.44} & \mbox{for}~L_\gamma > L_* \\
\end{array} \right.,
\end{eqnarray}
where $L_*=10^{52.5}~\mbox{erg}~\mbox{s}^{-1}$.
The minimum Luminosity is $10^{50}~\mbox{erg}~\mbox{s}^{-1}$.
The remaining parameters are $f_{\rm CR}$ and $\Gamma$.
We adopt four sets of those parameters as summarized in
Table~\ref{tab1} and obtain the average UHECR spectrum
per burst by integrating the function over the luminosity function
and using the relation of Eq. (\ref{eq:ghirel}).

\begin{table}[b]
\caption{\label{tab1}%
Model parameters.
}
\begin{ruledtabular}
\begin{tabular}{lcccc}
\textrm{Model}&
\textrm{A}&
\textrm{B}&
\textrm{C}&
\textrm{D}\\
\colrule
$f_{\rm CR}$ & 10 & 10 & \textrm{U.M.}\footnote{Universal CR luminosity
model expressed in Eq. (\ref{eq:fcr})} & \textrm{U.M.} \\
$\Gamma$ & 300 & $72.1 L_{52}^{0.49}$ & 300 & $72.1 L_{52}^{0.49}$ \\
\colrule
\textrm{LLC}\footnote{The UHECR contribution from GRBs with $L\leq L_*$
at $10^{18.5}$ eV (low luminosity contribution).}
& 30.0\% & 45.8\% & 92.3\% & 100\% \\
\end{tabular}
\end{ruledtabular}
\end{table}

\begin{figure}[t]
\includegraphics[width=0.45\textwidth]{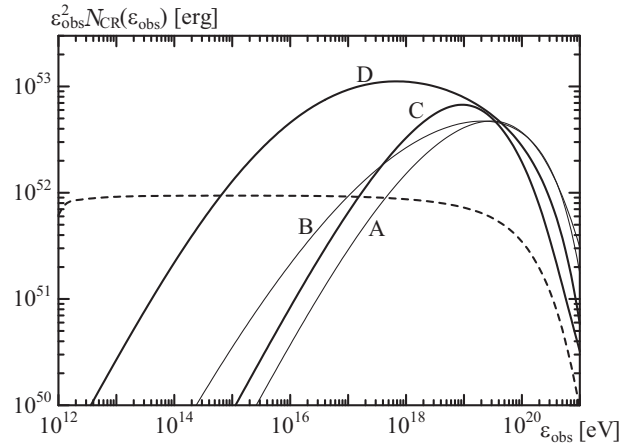}
\caption{\label{fig:avg} The average UHECR spectra per burst
for the parameter sets shown in Table~\ref{tab1}.
The thin lines are for the models A and B,
while the thick lines are for the models C and D.
The dashed line is the average UHECR spectrum for
the shock acceleration model adopted by \citet{asa14},
in which $f_{\rm CR}=10$, $f_B=0.1$, and $\Gamma_{300}=1$
with the same luminosity function.}
\end{figure}

In the model A, we adopt constant parameters $f_{\rm CR}=10$
and $\Gamma=300$ irrespective of the luminosity.
As shown in Fig. \ref{fig:avg}, unlike the shock acceleration model in
the work by \citet{asa14},
the CRs are narrowly distributed in the highest energy region.
Owing to the hard spectrum, the spectral peak at $\sim 10^{20}$ eV
becomes higher than that for the shock model.
To take into account the diversity of $\Gamma$,
we also adopt an empirical relation between $\Gamma$ and $L_\gamma$
expressed by $\Gamma=72.1 L_{52}^{0.49}$, which was adopted by \citet{He12}
based on the results of \citet{Ghi12}.
This relation leads to a slightly softer spectrum, as shown
by the model B in Fig. \ref{fig:avg}.

In models A and B, we adopted a common value of $f_{\rm CR}$,
irrespective of $L_\gamma$.  However, from the viewpoint of the energy budget,
bright GRBs may not have a large margin for UHECRs.
We may expect some correlation between $L_\gamma$ and $f_{\rm CR}$, 
similarly to the several empirical relations in GRBs \citep[e.g., Ref.][]{Ghi12}.

Although we have no information about the origin of the $L_\gamma$-$f_{\rm CR}$ 
relation, here we can adopt a simple model as a test case to demonstrate the possible 
broadening of the UHECR spectrum. For this, we assume that the total luminosity is 
nearly constant irrespective of $L_\gamma$.  In this model, the variety of the photon 
luminosity is attributed to the variety of the photon emission efficiency.
Normalizing at $L_\gamma=L_*$, we set $L_{\rm tot}= L_\gamma+L_B+L_{\rm CR}=
(1+f_B+f_{\rm CR}) L_\gamma=\max(10 L_*,(2+f_B) L_\gamma)$, where $L_B$ stands for 
magnetic field luminosity.  Choosing here a typical total jet luminosity as 
$L_{\rm tot}=10^{53.5}~\mbox{erg}~\mbox{s}^{-1}$, with $f_B=1$, this is expressed as
\begin{eqnarray}
f_{\rm CR}= \left\{ \begin{array}{ll}
10\frac{L_*}{L_\gamma}-2 & \mbox{for}~L_\gamma \leq (10/3) L_* \\
1 & \mbox{for}~L_\gamma > (10/3) L_* \\
\end{array} \right. .
\label{eq:fcr}
\end{eqnarray}
In this model, while the total luminosity is kept in check, the relatively 
low-luminosity GRBs, which dominate the number fraction of GRBs, are the dominant 
sources of UHECRs.  The fractional contribution of GRBs with $L \leq L_*$
to the UHECRs at $10^{18.5}$ eV for each model is shown in Table~\ref{tab1}.

While we have fixed the Lorentz factor as $\Gamma=300$ for the model C,
the relation $\Gamma=72.1 L_{52}^{0.49}$ is adopted in the model D.
As shown in Fig. \ref{fig:avg}, this optimistic $L_\gamma$-$f_{\rm CR}$ relation
provides a lower peak energy and a broader shape
for the average UHECR spectrum, especially for the model D.
Unlike in the shock acceleration model, the spectral shape
is not a simple power law, but shows a curved feature.

\section{UHECRs at the earth}

For the cosmic-ray propagation, we adopt the same method as \citet{asa14}. 
We calculate the comoving density of UHECRs $n_{\rm CR}$ taking into account 
the cooling effects due to the adiabatic cosmological expansion,
photomeson production, and Bethe-Heitler pair production
with the extra galactic background light model by \citet{kne04}.
UHECRs are injected at a rate according to \citet{wan10},
$R_{\rm GRB}(z) \propto (1+z)^{2.1}$ for $z \leq 3.0$
and $\propto (1+z)^{-1.4}$ for $z>3.0$
with the average spectrum obtained in the previous section.
The local rate is taken as $R_{\rm GRB}(0)=1.3~\mbox{Gpc}^{-3}~\mbox{yr}^{-1}$.
Assuming the standard cosmology, the integral over the redshift is performed
with the differential transformation
\begin{eqnarray}
\frac{dt}{dz}=-\frac{1}{(1+z) H_0 \sqrt{\Omega (1+z)^3+\Lambda}},
\end{eqnarray}
where $\Omega=0.3$, $\Lambda=0.7$, and $H_0=70~\mbox{km}~\mbox{s}^{-1}~\mbox{Mpc}^{-1}$.

\begin{figure}[t]
\includegraphics[width=0.45\textwidth]{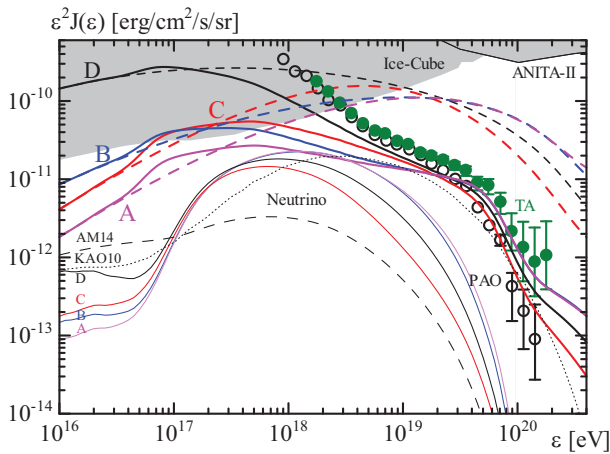}
\caption{\label{fig:res} The diffuse UHECR spectra for models A--D
(thick solid lines). The thick dashed lines are spectra
neglecting the effects of photomeson production and Bethe-Heitler
pair production.
The observed data for the UHECR intensities
are taken from \citet{sch13} for the Pierre Auger Observatory (open circles)
and \citet{Abu13} for the Telescope Array (green filled circles).
The thin lines show the all-flavor cosmogenic neutrino intensities
for the models A--D,
which are below the upper limits (gray shaded area) by IceCube
taken from \citet{hei15} based on \citet{ish15},
and ANITA-II \citep{gor12}.
For comparison,
we also plot the model spectra of the cosmogenic neutrinos
by \citet{kot10} (thin dotted line, denoted as KAO10)
and prompt plus cosmogenic neutrinos by
\citet{asa14} (thin dashed line, denoted as AM14).}
\end{figure}

The resultant UHECR intensities $J_{\rm CR}=c n_{\rm CR}/(4 \pi)$
for models A--D are presented in Fig. \ref{fig:res}.
While we have tested the two extreme models for $f_{\rm CR}$,
all the models seem to agree with the flux above $\sim 10^{19}$ eV.
At the ankle point ($10^{18.5}$ eV), model D is the one which
most closely agrees with the observations.
Depending on the cutoff shape of the spectral component
below the ankle point, all the other models are also
within the permissible range.
Therefore, if the typical cosmic-ray luminosity is $10^{53.5}~\mbox{erg}~\mbox{s}^{-1}$
as assumed in all the models, the stochastic acceleration models can explain the 
UHECR observations above the ankle, unless $f_B \ll 1$ or $\zeta \ll 1$.
The curved shape of the intrinsic UHECR spectra in our model does not induce
a significant difficulty on the final diffuse UHECR spectrum.

The effect of the intergalactic magnetic field (IGMF) on cosmic-ray
propagation is omitted in Fig. \ref{fig:res}. The cosmic-ray
flux below $\sim 10^{17}$ eV should be suppressed because of the
magnetic confinement near the sources \citep{ber07,mol13}.
However, the production rate of the secondary neutrinos
during the CR propagation,
namely, the cosmogenic (GZK) neutrinos, may not be largely affected by
the IGMF.
The intensity of the GZK neutrinos
is well below the observational upper limits.
While the neutrino upper limit gradually bends the proton dip model \citep{hei15},
our model belongs to the so-called ankle transition model.
Judging from the spectral shape of the IceCube upper limit,
the energy of the first GZK neutrinos to be expected in future detections
will be in the $10^{17.5}$--$10^{18}$ eV range.
In this energy range, all our models predict a similar intensity
because the intrinsic UHECR intensities (dashed lines in Fig. \ref{fig:res})
at $10^{19}$ eV (the typical energy of the parent protons for such neutrinos)
are also close to each other.
As a representative model characteristic of previous studies,
we also plot the GZK neutrino spectrum for the ankle transition model
by \citet{kot10} (WW model), in which UHECRs are
injected with a power-law spectrum of $p=2.1$
and a cutoff energy of $10^{20.5}$ eV following the star formation
rate derived in \citet{hop06}.
The total neutrino spectrum (prompt plus cosmogenic)
based on the shock acceleration
in GRBs by \citet{asa14} (the injection spectrum is shown in Fig. \ref{fig:avg}),
in which the UHECR intensity at the ankle energy is not reproduced, is also shown.
Our models here show the hardest spectra at $10^{17.5}$ eV,
compared to the previous models.

\section{Acceleration Site and Possible Neutrino Emission}

The radius $R$ of the UHECR acceleration site was not specified in the previous section.
The parameter relations adopted in this paper imply 
$E_\gamma=2 \times 10^{52}$ erg and $\Gamma=127$ for $L_\gamma=L_*$,
and the total jet energy $E_{\rm tot}$ is larger than $f_{\rm CR} E_\gamma$.
For $E_{\rm tot}=10^{53.5}$ erg, the jet starts to decelerate at the  radius
\begin{eqnarray}
R_{\rm dec}&=&\left( \frac{3 E_{\rm tot}}{4 \pi n m_{\rm p} c^2 \Gamma^2}
\right)^{1/3} \nonumber \\
&\simeq& 1.46 \times 10^{17} n_0
\left( \frac{E_{\rm tot}}{10^{53.5}~\mbox{erg}}
\right)^{1/3} \left( \frac{\Gamma}{127} \right)^{-2/3} \mbox{cm}, \nonumber \\ 
\end{eqnarray}
where the density of the interstellar medium is $n=n_0~\mbox{cm}^{-3}$.
The UHECR acceleration site may be around or inside this radius.
If the prompt gamma-ray photons are emitted from an inner radius
prior to the UHECR acceleration at the outer radius,
the photons may already have escaped at the onset time of the UHECR acceleration.
In this case, the cooling effect due to photopion production
on the UHECR spectrum can be neglected, and neutrino emission
is not expected.

\begin{figure}[t]
\includegraphics[width=0.45\textwidth]{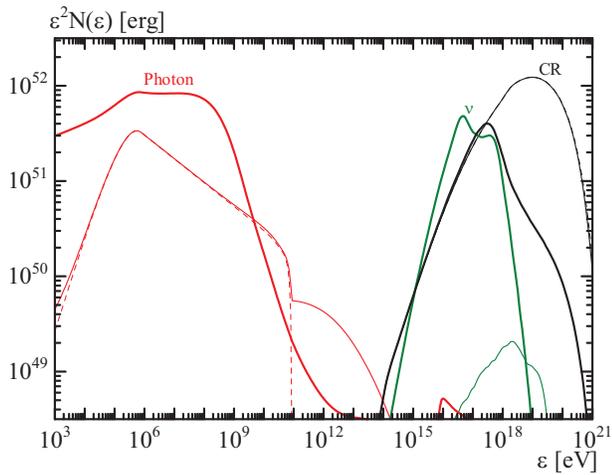}
\caption{\label{fig:sec} The final photon (red),
cosmic-ray (black), and neutrino (green) spectra
from a GRB
with $E_\gamma=2\times 10^{52}$ erg and $\Gamma=127$.
The assumed radii of the UHECR acceleration site
are $10^{15}$ cm (thick line), $10^{16}$ cm (thin line),
and $10^{17}$ cm (dashed line), respectively.
The dashed lines for the photon and cosmic ray mostly overlap
with the thin lines.
The photon spectrum for $10^{17}$ cm is almost the input shape
of the Band function.
The dashed line for the neutrino is far below the plot range
of this figure.}
\end{figure}

However, when the duration of the prompt emission $\Delta T$ is longer than
$R/(c \Gamma^2)\simeq 21 R_{16} (\Gamma/127)^{-2}$ s,
some fraction of the gamma-ray photons may 
still be in the acceleration region.
This would lead to a delayed onset of the neutrino emission
triggered by the $p \gamma$ collisions, which may be a signature of
the different radii of the UHECR acceleration and the prompt emission.
If the acceleration site radius is larger than
$c \Delta T \Gamma^2 \simeq 4.8 \times 10^{15} (\Gamma/127)^2
(\Delta T/10~\mbox{s})$ cm, the parameter $f_B$ may be
different from the value at the photon emission site.
In this case, the volume expansion may reduce the value as $f_B \propto R^{-1}$.
Although we consider the case $R>c \Delta T \Gamma^2$ while keeping $f_B=1$ below,
the modification of $f_B$ affects only the maximum energy
in Eq. (\ref{emax2}), as $\varepsilon_{\rm max} \propto R^{-1/2}$.
The neutrino production for such a large $R$ is inefficient,
irrespective of $f_B$, as discussed below.

Assuming the Band function for the prompt gamma-ray spectrum
(generic peak energy of $570$ keV, low- and high-energy 
photon indices of $-1$ and $-2.25$, respectively) with
an average photon density $E_\gamma/(4 \pi R^3)$,
which is the upper limit in this model
(the escape fraction of the prompt photons before the onset of
the UHECR acceleration is assumed to be zero),
we simulate the hadronic cascade with the same method as that in \citet{asa14}.
The parameter $f_{\rm CR}$ is given by Eq. (\ref{eq:fcr}) with $L_\gamma=L_*$.
As shown in Fig. \ref{fig:sec}, for $R=10^{17}$ cm and $10^{16}$ cm,
the cooling effect on UHECRs is negligible.
For $R=10^{15}$ cm, UHECRs lose their energies via photopion production,
and secondary gamma rays overwhelm the primary gamma rays.
Therefore, the UHECR acceleration at $R \lesssim 10^{15}$ cm
in the maximum prompt photon field has to be rejected for
this parameter set. An alternative option to suppress the neutrino flux
at radii $R=10^{15}$ cm
is to adopt a higher $\Gamma \gtrsim 1000$ as an average value.
For allowed radii such as $R=10^{16}$ cm or larger
in the case of $\Gamma=127$,
the neutrino fluence becomes much lower than the photon fluence,
so that we do not expect neutrino detection by IceCube. 
Also, from the usual branching ratio between charged and neutral pion
production, the corresponding high-energy gamma rays produced will be
at a comparable level to that of neutrinos. As discussed by various authors
(e.g. Refs. \citep{Bechtol+16nusfg,Tamborra+14sfgnugam,Murase+13pev}), sources which
could reproduce the flux of the extragalactic diffuse
PeV neutrinos detected with IceCube \citep{aar15} are at risk of violating the 
diffuse 1-800 GeV gamma-ray background seen by {\it Fermi}. Thus, since in our case
the neutrino flux is well below the IceCube limits, also the related diffuse 
GeV gamma-ray background is expected to be well below the Fermi limits.

Although neutrino emission from the UHECR production site is not 
expected to be significant, as discussed above, if the prompt gamma-ray 
emission arises from a dissipative photosphere, this may also result in 
neutrino emission \citep[e.g., Refs.][]{gao12,zha-kum13}.
The neutrino upper limits in the PeV energy range \citep{mur13} already
exclude $f_{\rm CR} \gtrsim 10$ at the photosphere, which may imply,
as we have assumed here, that UHECR acceleration is suppressed
at such small radii. Neutrinos of 10--100 GeV may also be produced 
below the photosphere as a result of $p$-$n$ or $p$-$p$ collisions
\citep[e.g., Refs.][]{bah00,asa13,bar13,mur13}. Such low-energy neutrino 
emission, however, is not observationally constrained yet.

Electrons can also be accelerated by the same stochastic process caused by 
turbulence. However, the electron acceleration time scale $\sim 0.3 \xi_{0.1} R/(c \Gamma)$
in this model is so long that the synchrotron cooling effect prevents
the electron acceleration
[the cooling time scale $6 \pi m_{\rm e} c/(\sigma_{\rm T} B^2 \gamma)$ results in a 
maximum Lorentz factor $7.5 f_B^{-1} \xi_{0.1}^{-1} R_{17} (L_*/L_\gamma) (\Gamma/127)^3$].
The photon field emitted by electrons can therefore be neglected
as target photons for $p \gamma$ collisions.

If the turbulence responsible for the UHECR acceleration arises in shocks 
like the forward/reverse shock with the Rayleigh-Taylor instability \citep{duf13}
or in internal shocks with the Richtmyer-Meshkov instability \citep{ino11},
a first-order Fermi acceleration may also act upon the electrons.
However, the Rayleigh-Taylor fingers arising from the instability
can disrupt the smooth laminar shock structure needed for the first-order 
Fermi acceleration, and similar effects may also be associated with the 
Richtmyer-Meshkov instability.  The photon emission from the forward shock 
afterglow does not provide large cooling effect on UHECRs \citep{mur07}.
Also, the x-ray flares \citep[see e.g., Ref.][]{chi10} in the early afterglow phase
may be a signature of late internal shocks \citep{fan05}, which may drive turbulence.
Unless $R \lesssim 10^{15}$ cm (corresponding to a few seconds for the flare duration),
the photopion production due to x-ray flares is not efficient enough to cool the UHECRs
\citep{mur06}. Therefore, the results shown in Fig. \ref{fig:sec} are unlikely to be
significantly altered by the photons from the shock-accelerated electrons.

\section{Conclusion}

We propose a possible model of the UHECR production by the 
stochastic proton acceleration via turbulence in the GRB jets.
The UHECR spectrum at injection is harder than in previous models
and shows a curved feature, which does not conflict with the 
observed UHECR spectral shape, its presence being felt mainly
above the ankle.
The required typical cosmic-ray luminosity is $\sim 10^{53.5}~
\mbox{erg}~\mbox{s}^{-1}$, which is moderate compared to previous
GRB UHECR models. An overly luminous secondary gamma-ray/neutrino emission
initiated by photopion production is avoided because the acceleration 
site is expected to be well outside the photon emission radius.
A predicted hard spectrum of GZK neutrinos in the $10^{17}$--$10^{18}$ eV 
range can be a clue to constraining the parent UHECR spectrum.

The UHECR spectrum at injection is very sensitive to the model parameters, 
which are uncertain and may have a substantial dispersion.
Especially the $L_\gamma$-$f_{\rm CR}$ or $L_\gamma$-$\Gamma$ relations 
are not well defined. 
Depending on those parameters, the dominant UHECR contribution
may come from the relatively low-luminosity GRBs
($L_\gamma<10^{52.5}~\mbox{erg}~\mbox{s}^{-1}$) or vice versa.
Although we cannot, so far, predict a quantitatively precise UHECR spectrum,
the possibility of a hard spectrum such as discussed in this paper appears
to be an attractive idea for overcoming the difficulties in the GRB UHECR hypothesis.

\begin{acknowledgments}
First, we thank the anonymous referee for very useful advice
and careful inspection.
We wish to acknowledge T. Terasawa and Y. Teraki for valuable discussion.
This work is partially supported by the Grant-in-Aid for Scientific Research
No. 16K05291 from the MEXT of Japan (K.A.) and by NASA Grant No. NNX13AH50G (P.M.).
\end{acknowledgments}

\bibliography{apssamp}

\end{document}